\documentclass[prl,twocolumn,nofootinbib,amsfonts]{revtex4}
\usepackage{graphicx,bm,epsfig,amsmath}
\newcommand{\be}{\begin{equation}}
\newcommand{\ee}{\end{equation}}
\newcommand{\dH}{\dot{H}}
\newcommand{\da}{\dot{a}}
\newcommand{\ddH}{\ddot{H}}
\newcommand{\dda}{\ddot{a}}

\begin{document}

\title{An extension of the cosmological standard model \\ with a bounded
  Hubble expansion rate}
\author{J.L. Cort\'es}
\email{cortes, indurain@unizar.es}
\affiliation{Departamento de F\'{\i}sica Te\'orica,
Universidad de Zaragoza, Zaragoza 50009, Spain}
\author{J. Indur\'ain}
\email{cortes, indurain@unizar.es}
\affiliation{Departamento de F\'{\i}sica Te\'orica,
Universidad de Zaragoza, Zaragoza 50009, Spain}

\begin{abstract}
The possibility of having an extension of the cosmological standard
model with a Hubble expansion rate $H$ constrained to a finite
interval is considered. Two periods of accelerated expansion arise
naturally when the Hubble expansion rate approaches to the two
limiting values. The new description of the history of the universe is
confronted with cosmological data and with several theoretical ideas
going beyond the standard cosmological model.
\end{abstract}

\maketitle

\section{Introduction}
According to General Relativity, if the universe is filled with the particles
of the Standard Model of particle physics, gravity should lead to a
deceleration of the expansion of the universe. However, in 1998 two
independent evidences of present accelerated expansion were presented
\cite{Riess1,Perlmutter1} and later confirmed by different
observations \cite{KOBE, WMAP,LSS}. On the other hand, measurements of large
scale structure \cite{infLSS} and CMB anisotropy \cite{KOBE} also
indicate that the universe evolved through a period of early accelerated
expansion (inflation).

There is no compelling explanation for any of these cosmic accelerations,
but many intriguing ideas are being explored. In the case of inflation,
the origin of the accelerated expansion can be either a modification of
gravity at small scales \cite{hemg} or a coupling of the expansion of
the universe to the progress of phase transitions \cite{inflation}.
In the case of the present accelerated expansion these ideas can be classified
into three main groups: new exotic sources of the gravitational field with
large negative pressure \cite{DE} (Dark Energy), modifications of
gravity at large scales \cite{lemg} and rejection of the spatial
homogeneity as a good approximation in the description of the present
universe \cite{DEwoDE}.

Different models (none of them compelling) of the source responsible
for each of the two periods of accelerated expansion have been
considered. Einstein equations admit a cosmological constant
$\Lambda$, which can be realized as the stress-energy tensor of empty
space. This $\Lambda$, together with Cold Dark Matter, Standard Model
particles and General Relativity, form the current cosmological model
($\Lambda$CDM). However, quantum field theory predicts a value of
$\Lambda$ which is 120 orders of magnitude higher than observed.
Supersymmetry can lower this value 60 orders of magnitude, which is
still ridiculous \cite{Weinberg}.
In order to solve this paradox, dynamical Dark Energy models have been proposed.

This has also lead to explore the possibility that cosmic acceleration arises
from new gravitational physics. Also here several alternative
modifications of the Einstein-Hilbert action at large and small
curvatures \cite{fR, Carroll1, Soti, HS, Odin, Odin2, Odin8}, or even higher
dimensional models \cite{Deffayet, DGP}, producing an accelerated
expansion have been identified.
All these analysis include an \textit{ad hoc} restriction to actions involving simple
functions of the scalar curvature and/or the Gauss-Bonnet tensor.
This discussion is sufficient to establish the point that cosmic acceleration can
be made compatible with a standard source for the gravitational field, but it is
convenient to consider a more general framework in order to make a systematic
analysis of the cosmological effects of a modification of general relativity.

In this paper we parametrize the evolution of the universe (considered isotropic and
homogeneous) with the Hubble parameter $H$. One finds that it is possible to restrict
the domain of $H$ to a bounded interval. This restriction naturally
produces an accelerated expansion when the Hubble expansion rate
approaches any of the two edges of the interval.
Therefore, we find a new way to incorporate two periods
of cosmic acceleration produced by a modification of general
relativity. But the dependence on the two limiting values
of $H$ can be chosen independently. One could even consider a Hubble
expansion rate constrained to a semi-infinite interval with a unique
period of accelerated expansion. In this sense the aim to
have a unified explanation of both periods of accelerated expansion is
only partially achieved. This simple phenomenological approach to the problem
of accelerated expansion in cosmology proves to be equivalent at
the homogeneous level to other descriptions based on modifications
of the Einstein-Hilbert action or the introduction of exotic components
in the matter Lagrangian.

In the next section we will review how a general modification of the gravitational
action leads to a generalized first Friedman equation at the homogeneous level. In
the third section we will present a specific model (ACM) based on the simplest
way to implement a bounded interval of $H$. In the fourth section we will contrast
the predictions of ACM for the present acceleration with astrophysical observations.
In the fifth section we will show that it is always possible to find modified
gravitational actions which lead to a given generalized first Friedman equation, and
present some simple examples. In the sixth section we will show that it is also always
possible to find Dark Energy models which are equivalent to a given generalized
first Friedman equation, and present some simple examples. The last section is devoted
to summary and conclusions.

\section{Action of the cosmological standard model extension}

The spatial homogeneity and isotropy allow to reduce the
gravitational system to a mechanical system with two
variables $a(t)$, $N(t)$ which parametrize the Robertson-Walker
geometry
\be
ds^2 \,=\, N^2(t) dt^2 - a^2(t) \Delta_{ij} dx^i dx^j
\ee
\be
\Delta_{ij} \,=\, \delta_{ij} + \frac{k x_i x_j}{1 - k{\bf x}^2} \, .
\ee
Invariance under parameterizations of the time variable
imply that the invariant time differential $N(t)dt$ must be used. Also
a rescaling of the spatial variables
$x^i\rightarrow\lambda x^i$ together with $a(t)\rightarrow\lambda^{-1}
a(t)$ and $k\rightarrow\lambda^{-2} k$ is a symmetry that must be kept
in the Lagrangian. The action of the reduced homogeneous gravitational
system can be then written as
\be
I_g \,=\, \int dt N \, L\left(\frac{k}{a^2}, H,
\frac{1}{N}\frac{dH}{dt},
\frac{1}{N}\frac{d}{dt}\left(\frac{1}{N}\frac{dH}{dt}\right),
...\right) \, ,
\label{Ig}
\ee
with $H \,=\, \frac{1}{aN}\frac{da}{dt}$.

If we keep the standard definition of the gravitational coupling and
the density ($\rho$) and pressure ($p$) of a cosmological homogeneous
and isotropic fluid as a source of the gravitational field, we have
the equations of the reduced system
\be
\left(\frac{8\pi G_N}{3}\right) \rho \,=\, - \frac{1}{a^3} \frac{\delta
  I_g}{\delta N(t)}
\ee
\be
- \left(8\pi G_N\right) p \,=\, - \frac{1}{N a^2} \frac{\delta
  I_g}{\delta a(t)}\, .
\ee
It is possible to choose a new time coordinate $t^{'}$ such that
\be
\frac{dt^{'}}{dt} \,=\, N(t)\, .
\ee
This is equivalent to set $N=1$ in the action (\ref{Ig}) of the gravitational
system and the evolution equations of the cosmological model reduce to a
set of equations for the scale factor $a(t)$. If we introduce the
notation
\be
H^{(i)} \,=\, \left(\frac{d}{dt}\right)^i \, H
\ee
\be
\left(\delta_i L\right)^{(j)} \,=\, \left(\frac{d}{dt}\right)^j
\left[a^3 \partial_i L\right]\, ,
\ee
with $\partial_i L$ denoting the partial derivatives of the Lagrangian
as a function of the variables $(k/a^2, H, H^{(1)}, H^{(2)}, ...)$ we
have
\be
\left(\frac{8\pi G_N}{3}\right) \rho \,=\, - L +
\sum_{i=2}^{\infty} \sum_{j=0}^{i-2} \frac{ (-1)^{i-j}}{a^3} H^{(j)} \left(\delta_i
L\right)^{(i-j-2)}
\label{rho}
\ee
\be
- \left(8\pi G_N\right) p \,=\, - 3 L + \frac{2 k}{a^2} \partial_1 L +
 \sum_{i=2}^{\infty} \frac{(-1)^i}{a^3} \left(\delta_i
L\right)^{(i-1)}   \, .
\label{p}
\ee
In the homogeneous and isotropic approximation, the vanishing of the
covariant divergence of the energy-momentum tensor leads to the
continuity equation
\be
\frac{d}{dt} \left(\rho a^3\right) \,=\, - p \frac{d}{dt} a^3
\label{cont_eq}\, .
\ee
In the radiation dominated era one has
\be
\rho \,=\ 3 p \,=\, \frac{\sigma}{a^4}
\label{rad}\, ,
\ee
where $\sigma$ is a constant parameterizing the general
solution of the continuity equation. In a period dominated by matter
one has a pressureless fluid and then
\be
p \,=\, 0 \, ,{\hskip 2cm} \rho \,=\, \frac{\eta}{a^3}\, ,
\ee
with constant $\eta$. When these
expressions for the energy density and pressure are plugged in
(\ref{rho}-\ref{p}), one ends up with two compatible differential equations for
the scale factor $a(t)$ which describe the evolution of the
universe.  From now on we will use the more common notation
$\frac{dH}{dt}\equiv\dot{H}$,...

\section{The Asymptotic Cosmological Model}
Let us assume that the Lagrangian L of the gravitational system is
such that the evolution equations (\ref{rho}-\ref{p}) admit a solution
such that $a(t)>0$ (absence of singularities), $\da >0$ (perpetual
expansion) and $\dH<0$. In that case, one has a different value of
the scale factor $a$ and the Hubble rate $H$ at each time and then one
has a one to one correspondence between the scale factor and the
Hubble rate. Since the continuity equation (together with the equation
of state) gives a relation between the scale factor and the density,
one can describe a solution of the evolution equations of the
generalized cosmological model through a relation between the energy
density and the Hubble rate, i.e. through a generalized first Friedman
equation
\be
\left(\frac{8\pi G_N}{3}\right) \rho \,=\, g(H)\, ,
\label{rho_g}
\ee
with $g$ a smooth function which parametrizes the different algebraic
relations corresponding to different solutions of different
cosmological models. Each choice for the function $g(H)$ defines a
phenomenological description of a cosmological model. Then one
can take it as a starting point trying to translate any observation
into a partial information on the function $g(H)$ which parametrizes
the cosmological model.

Eq. (\ref{rho_g}) is all one needs in order to reconstruct the
evolution of the universe at the homogeneous level. The generalized
second Friedman equation is
obtained by using the continuity equation (\ref{cont_eq}) and the
expression for $\rho$ as a function of $H$. One has
\be
- \left(8\pi G_N\right) p \,=\, 3 g(H) + \frac{g^{'}(H)}{H} \dH
\label{pressure}\, .
\ee
In the matter dominated era one has
\be
\frac{\dH}{H^2} \,=\, - 3 \frac{g(H)}{H g^{'}(H)}
\label{matter}
\ee
and then the assumed properties of the solution for the evolution
equations require the consistency conditions
\be
g(H) > 0 \, ,{\hskip 1cm} g^{'}(H) >0\, .
\label{gcons}
\ee
In the period dominated by radiation one has
\be
\frac{\dH}{H^2} \,=\, - 4 \frac{g(H)}{H g^{'}(H)}
\label{radiation}
\ee
instead of (\ref{matter}) and the same consistency conditions
(\ref{gcons}) for the function $g(H)$ which defines the generalized
first Friedman equation.

We introduce now a phenomenological cosmological model
defined by the condition that the Hubble rate has an upper bound $H_+$
and a lower bound $H_-$. This can be implemented through a function
$g(H)$ going to infinity when $H$ approaches $H_+$ and going to
zero when $H$ approaches $H_-$. We will also assume that there is an interval
of $H$ in which the behavior of the energy density with the Hubble parameter is,
to a good approximation, scale-free i.e. $g(H)\propto H^2$. The source
of the gravitational field will be a homogeneous and isotropic fluid
composed of relativistic and non-relativistic particles.
The Cosmological Standard Model without curvature is recovered in the
limit $H_-/H\rightarrow 0$ and $H_+/H\rightarrow\infty$ which is a
good approximation for the period of decelerated expansion.

Notice that this interpretation is independent of the underlying
theory of gravitation.
The Hubble parameter can be used to parametrize the history of
universe as long as $\dH\neq0\,\forall \,t$. The total density can be
thus expressed as a function of $H$. If $H$ is bounded, then $\rho(H)$
will have a pole at $H=H_+$ and a zero at $H=H_-$. Far from these scales,
the behavior of $\rho(H)$ can be assumed to be approximately scale-free.

Under these conditions, we can parametrize the dependence of the
cosmological model on the lower bound $H_-$ by
\be
g(H) \,=\, H^2 h_{-}\left(\frac{H_-^2}{H^2}\right)
\label{g-}
\ee
and similarly for the dependence on the upper bound $H_+$
\be
g(H) \,=\, \frac{H^2}{
    h_{+}\left(\frac{H^2}{H_+^2}\right)}
\label{g+}\, ,
\ee
where the two functions $h_{\pm}$ satisfy the conditions
\be
\lim_{x\to 0} h_{\pm} (x) \,=\, \beta^{\pm 1}\, , {\hskip 1cm}
\lim_{x\to 1} h_{\pm} (x) \,=\, 0\, .
\ee
$\beta$ is a constant allowed in principle by dimensional
arguments. If $\beta \neq 1$ then it can be moved to the lhs of the
Friedman equation, turning $G_N\rightarrow\beta G_N$, and can be
interpreted as the ratio between an effective cosmological value of
the gravitational coupling and the value measured with local
tests. But $\beta\neq 1$ would be in conflict with Nucleosynthesis,
through the relic abundances of $^4 He$ and other heavy elements (for
3 neutrino species) \cite{Mukh}, so we set $\beta= 1$ . Therefore
\be
\lim_{x\to 0} h_{\pm} (x) \,=\, 1 \, ,{\hskip 1cm}
\lim_{x\to 1} h_{\pm} (x) \,=\, 0\, .
\label{hpm}
\ee

The consistency conditions (\ref{gcons}) result in
\be
h_{\pm} (x) > 0 \, ,{\hskip 1cm} h_{\pm} (x) > x h_{\pm}^{'}(x)
\label{hcons}
\ee
for the two functions $h_{\pm}$ defined in the interval $0<x<1$. Thus,
we can divide the cosmic evolution history into three periods. In the
earliest, relativistic particles dominate the energy density of the
universe and the generalized first Friedman equation shows a
dependence on the upper bound $H_+$. There is also a transition period
in which the effect of the bounds can be neglected and the rhs of the
first Friedman equation is scale-free; this period includes the
transition from a radiation dominated universe to a matter dominated
universe. In the third present period, non-relativistic particles
dominate the energy density of the universe but the dependence on the
lower bound $H_-$ must be accounted for in the generalized first
Friedman equation.

\indent From the definition of the Hubble parameter, one has
\be
\frac{\dH}{H^2} \,=\, -1 + \frac{a \dda}{\da^2}\, .
\ee
Then, in order to see if there is an accelerated or decelerated
expansion, one has to determine whether $\dH/H^2$ is greater or smaller
than $-1$.

In the period dominated by radiation one has
\be
\frac{\dH}{H^2} \,=\, - 2 \, \frac{1}{1 - \frac{x h_+^{'}(x)}{
    h_+(x)}}\, ,
\ee
with $x=H^2/H_{+}^2$, where we have used (\ref{g+}) assuming that only
the dependence on the upper bound ($H_{+}$) of the Hubble parameter is
relevant. A very simple choice for this dependence is given by
\be
g(H) \,=\, \frac{H^2}{\left(1 - \frac{H^2}{H_{+}^2}\right)^{\alpha_+}}
\label{alpha+}\, ,
\ee
with $\alpha_+$ a (positive) exponent which parametrizes the departure
from the standard cosmological model when the Hubble rate approaches
its upper bound. With this simple choice one has a transition from an
accelerated expansion for $H^2 > H_+^2/(1+\alpha_+)$ into a decelerated
expansion when $H^2 < H_+^2/(1+\alpha_+)$, which includes the domain of
validity of the standard cosmological model ($H\ll H_+$).

In the period dominated by matter (which corresponds to lower values
of the Hubble rate) we assume that only the dependence on the lower
bound ($H_-$) of the Hubble rate is relevant. Then one has
\be
\frac{\dH}{H^2} \,=\, - \frac{3}{2} \, \frac{1}{1 - \frac{x
    h_-^{'}(x)}{h_-(x)}}\, ,
\ee
with $x=H_{-}^2/H^2$. We can also consider a dependence on $H_-$
parametrized simply by an exponent $\alpha_-$
\be
g(H) \,=\, H^2 \, \left(1 - \frac{H_-^2}{H^2}\right)^{\alpha_-}
\label{alpha-}\, .
\ee
With this choice one has a transition from a decelerated expansion
for $H^2 > H_{-}^2(1+\alpha_-/2)$ leaving the domain of validity of
the standard cosmological model and entering into an accelerated
expansion when the Hubble rate approaches its lower bound for $H^2 <
H_{-}^2(1+\alpha_-/2)$.

The possibility of describing $\rho$ as a function of $H$ is independent
of the existence of spatial curvature $k$. However, in the kinematics of
observables in the expanding universe we do need to specify the value of
$k$. In the rest of the paper we will assume that the universe is flat
($k=0$), although the same analysis could be done for arbitrary $k$.

The properties of the expansion obtained in this simple example (a
period of decelerated expansion separating two periods of accelerated
expansion) are general to the class of phenomenological models with a
generalized first Friedman equation (\ref{rho_g}) with $g(H)$
satisfying the consistency conditions (\ref{gcons}) and a Hubble rate
constrained to a finite interval. The specific part of the example
defined by (\ref{alpha+},\ref{alpha-}) is the simple dependence on the
Hubble rate bounds and the values of the Hubble rate at the
transitions between the three periods of expansion. From now on we
will name this description the Asymptotic Cosmological Model (ACM).
With respect to the late accelerated expansion, ACM can be seen as a
generalization of $\Lambda$CDM, which can be recovered by setting
$\alpha_-=1$. It also includes an early period of exponential
expansion which can be seen as a phenomenological description of the
evolution of the universe at inflation in the homogeneous approximation.

\subsection{Horizon problem}
One can see that the horizon of a radiation-dominated universe can be
made arbitrarily large as a consequence of an upper bound on the
Hubble parameter and in this way one can understand the observed
isotropy of the cosmic microwave background at large angular scales.

Let us consider the effect of the modification of the cosmological
model on the calculation of the distance $d_h(t_f,t_i)$ of a source of
a light signal emitted at time $t_i$ and observed at time $t_f$
\be
d_h(t_f, t_i) \,=\, c \, a(t_f) \int_{t_i}^{t_f} \frac{dt}{a(t)}\, .
\ee
We have
\be
\frac{dt}{a} \,=\, \frac{da}{a^2 H}\,=\, \left(\frac{8 \pi G_N
  \sigma}{3}\right)^{-1/4} \frac{g^{'}(H) dH}{4H g(H)^{3/4}} \, ,
\ee
where in the first step we have used the definition of the Hubble
expansion rate $H$ and in the second step we have used the relation
between the scale factor $a$ and $H$ as given by
(\ref{rad}-\ref{rho_g}). We are considering both times $t_i$ and $t_f$
in the radiation dominated period.

The distance $d_h$ is then given by
\be
d_h(H_f, H_i) \,=\, \frac 1{4 g(H_f)^{1/4}} \int_{H_f}^{H_i} \frac{dH
  g^{'}(H)}{H g(H)^{3/4}} \label{dist}\, .
\ee
If $H_i$ is very close to $H_+$ (i.e. if we choose the time $t_i$ when
the light signal is emitted well inside the period of accelerated
expansion) then the integral is dominated by the
region around $H_i$ which is very close to $H_+$. Then one can
approximate in the integrand
\be
g(H) \approx \frac{H_+^2}{\left(1-\frac{H^2}{H_+^2}\right)^{\alpha_+}} \, .
\ee
On the other hand if the observation is made at a time $t_f$ within the
domain of validity of the cosmological standard model ($H_-^2\ll
H_f^2\ll H_+^2$) then the factor $g(H_f)^{1/2}$ in
front of the integral can be approximated by $H_f$ and then one has
\be
d_h(H_f, H_i) \approx \frac{\alpha_+}{4 \sqrt{H_f H_i}}
B_{\frac{H^2_i}{H^2_+}}(1/2,-\alpha/4) \, ,
\ee
where $B_z(m,n)$ is the incomplete Beta function, and it can be made
arbitrarily large by choosing $H_i$ sufficiently close
to $H_+$. In this way we see that a cosmological model with a finite
interval of variation for $H$ solves the horizon problem.

\section{Constraints of ACM by Observations}

In this section we will carry out a more technical analysis about how
the astrophysical observations constrain the parameter space of ACM in
the matter dominance period. This analysis is based on the use of
(assumed) standard candles, basically Type Ia Supernovae \cite{Riess2}
and CMB \cite{WMAP3y}. These observations constrain the parameter
space to confidence regions in which the combination
$\Omega_m\equiv(1-H_-^2/H_0^2)^{\alpha_-}$ is constrained to be around
one quarter.
The consideration of both Type Ia SNe and CMB together favor
$\alpha_->1.5$. The results of this analysis can be seen in
figures (FIG. 1-3).
A reader not interested in technical details might well skip this section.

We center our discussion of the experimental tests of the Asymptotic
Cosmological Model
in the late accelerated expansion produced when $H$ approaches its
lower bound $H_-$. The vast amount of supernovae data collected by
\cite{Riess2} and \cite{Astier}, the data from the SDSS  Baryon
Acoustic Oscillation \cite{Eisen}, the mismatch between total energy
density and total matter energy density seen at CMB anisotropies
\cite{WMAP3y} and the measurements of present local mass density by
2dF and SDSS \cite{LSS,Cole} compared with the measurements of $H_0$
from the HST Cepheids \cite{Cepheids} show that the universe undergoes
a surprising accelerated expansion at the present time.

We will firstly confront the model with the Supernovae Ia data from
Riess \textit{et al.} and SNLS collaboration.
The usefulness of the Supernovae data as a test of Dark Energy models
relies on the assumption that Type Ia SNe behave as standard candles,
i.e., they have a well defined environment-independent luminosity
$\mathcal{L}$ and spectrum. Therefore, we can use measured bolometric
flux $\mathcal{F}=\frac{\mathcal{L}}{4\pi d_L^2}$ and frequency to
determine luminosity distance $d_L$  and redshift $z$.
The luminosity distance is given now by
\be
\begin{array}{c}
d_L(z)=c(1+z)\int_0^z \frac{dz'}{H(z')}\\
=\frac{c(1+z)}{3}\int^{H(z)}_{H_0}\frac{g'(H)dH}{H g^{1/3}(H_0)g^{2/3}(H)}\, .
\end{array}
\ee
The computed value must be compared with the one obtained
experimentally from the measured extinction-corrected
distance moduli ($\mu_0=5 log_{10}(\frac{d_L}{1 Mpc})+25$) for each
SN. The SNe data have been compiled in references \cite{Riess2}, and
we have limited the lowest redshift at $cz<7000 km/s$ in order to
avoid a possible ``Hubble Bubble'' \cite{Jha,Zehavi}. Therefore our
sample consists on 182 SNe. We will determine the likelihood of the
parameters from a $\chi^2$ statistic,
\be
\chi^2(H_0,H_-,\alpha_-)=
\sum_i\frac{(\mu_{p,i}(z_i;H_0,H_-,\alpha_-)-\mu_{0,i})^2}{\sigma^2_{\mu_{0,i}}+\sigma^2_v}\,,
\ee
where $\sigma_v$ is the dispersion in supernova redshift due to
peculiar velocities  (we adopt $\parallel\mathbf{v_p}\parallel = 400
km/s$ in units of distance moduli), $\sigma_{\mu_{0,i}}$
is the uncertainty in the individual measured distance moduli
$\mu_{0,i}$, and $\mu_{p,i}$ is the value of $\mu_0$ at $z_i$
computed with a certain value of the set of parameters
${H_0,H_-,\alpha_-}$. This $\chi^2$ has been marginalized over
the nuisance parameter $H_0$ using the adaptive method in reference
\cite{Wang}. The resulting likelihood
distribution function $e^{-\chi^2/2}$ has been explored using Monte
Carlo Markov Chains. We get a best fit
of ACM at $\alpha_-=0.36$ and $\frac{H_-}{H_0}=0.95$, for which
$\chi^2=157.7$. In contrast, fixing $\alpha_-=1$, we get
$\Omega_\Lambda\equiv\frac{H_-^2}{H_0^2}=0.66$ and
$\chi^2=159.1$ for the best fit $\Lambda$CDM. The confidence regions
are shown in Fig. 1 (top).

\begin{figure}
 \centerline{\includegraphics[scale=0.6,
 width=9cm]{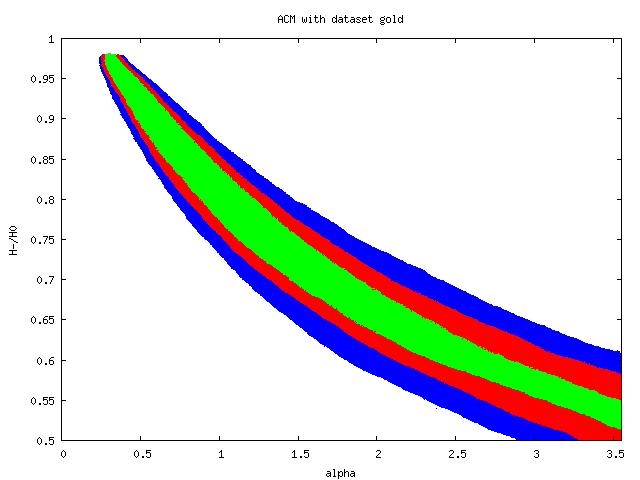}}
\end{figure}
\begin{figure}

\centerline{\includegraphics[scale=0.6,
 width=9cm]{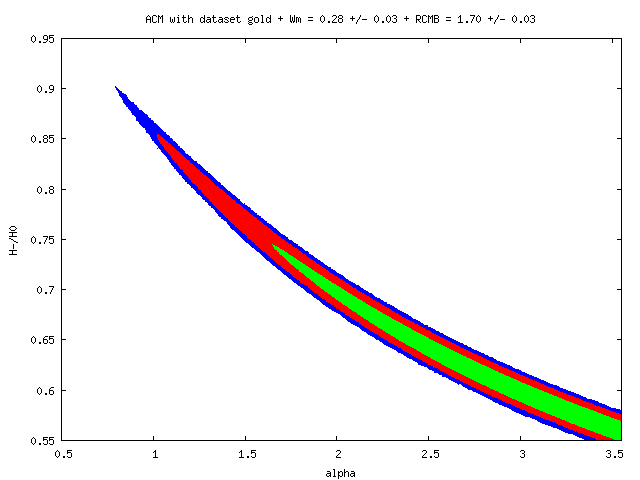}}

\label{confidence}
\caption{Confidence regions in parameter space of the Asymptotic
  Cosmological Model (ACM) with (bottom) and without (top) priors from
  CMB and estimations of present matter energy density at $1\sigma$
  (green), $2\sigma$ (red) and $3\sigma$ (blue). The $\Lambda$CDM is
  inside the 1$\sigma$ region if we consider no priors, but is moved
  outside the $2\sigma$ region when we confront the SNe data with CMB
  and estimations from present matter energy density. In this case a
  value $\alpha_-> 1.5$ is favored.}
\end{figure}

We can add new constraints for the model coming from measurements of
the present local matter energy density from the combination of 2dF
and SDSS with HST Cepheids, rendering $\Omega_m = 0.28 \pm 0.03$; and
the distance to the last scattering surface from WMAP, which leads to
$r_{CMB}\equiv\sqrt{\Omega_m}\int^{1089}_0 \frac{H_0
  dz'}{H(z')}=1.70\pm0.03$ \cite{Riess2}.
Including these priors we get a best fit of ACM at $\alpha_-=3.54$ and
$\frac{H_-}{H_0}=0.56$ (our simulation explored the region with
$\alpha_-<3.6$), for which $\chi^2=164.1$. In contrast, we get
$\Omega_\Lambda=0.73$ and $\chi^2=169.5$ for the best fit
$\Lambda$CDM, which is outside the $2\sigma$ confidence region shown
in Fig. 1 (bottom). The fits of the best fit $\Lambda$CDM and ACM
taking into account the priors to the SNe data are compared in Fig. 2.
The information which can be extracted from the data is limited. This
can be seen in Fig. 3, in which it is explicit that the data
constrain mainly the value of the present matter energy density,
$\Omega_m\equiv\frac{\rho_m}{\rho_C}=(1-\frac{H_-^2}{H_0^2})^{\alpha_-}$.

\begin{figure}
 \centerline{\includegraphics[scale=0.6, width=9cm]{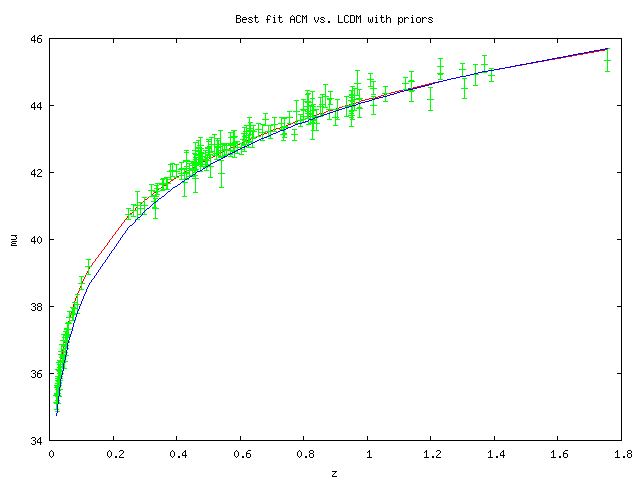}}
\label{bestfit}
\caption{Distance Moduli vs. Redshift comparison between the best
  fits of ACM ($H_0 = 66 \, Km/s MPc$, red) and $\Lambda$CDM ($H_0 =
  65 \, Km/s MPc$, blue) to the SNe data with priors. ACM fits clearly
  better the medium redshift SNe. }
\end{figure}

\begin{figure}
 \centerline{\includegraphics[scale=0.6, width=9cm]{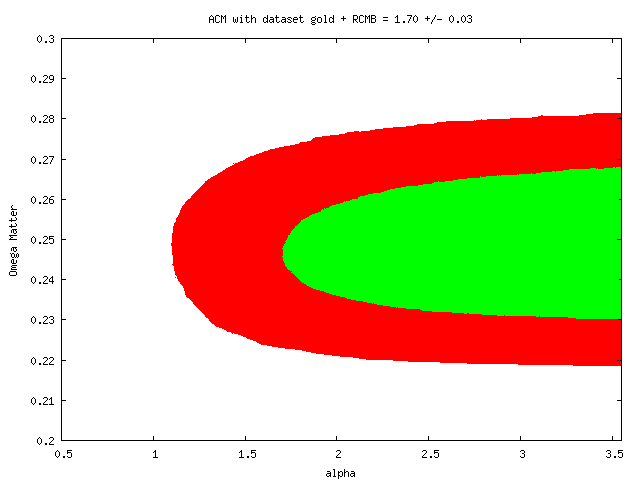}}
\label{omegabestfit}
\caption{$\Omega_m$ vs. $\alpha$ confidence regions with the CMB prior
  only. The confidence regions show $\Omega_m = 0.25 \pm 0.03$ and
  $\alpha > 1.15$ at $2\sigma$ level. }
\end{figure}

\section{Generalized First Friedman Equation and $f(R)$ gravity}

The extension of the cosmological model considered here could be
compared with recent works on a modification of gravity at large
or very short distances \cite{fR, Carroll1, Odin2, Soti, HS, Odin, Odin8}.
It has been shown that, by considering a correction to the
Einstein Hilbert action including positive and negative powers of the
scalar curvature, it is possible to reproduce an accelerated expansion
at large and small values of the curvature in the cosmological
model. Some difficulties to make these modifications of the
gravitational action compatible with the solar system tests of general
relativity have lead to consider a more general gravitational action,
including the possible scalars that one can construct with the
Riemann curvature tensor \cite{RR}, although the Gauss-Bonnet scalar is the only
combination which is free from ghosts and other pathologies. In fact, there
is no clear reason to restrict the extension of general relativity
in this way. Once one goes beyond the derivative expansion, one should
consider scalars that can be constructed with more than
two derivatives of the metric and then one does not have a good
justification to restrict in this way the modification of the
gravitational theory. It does not seem difficult to find an
appropriate function of the scalar curvature or the Gauss-Bonnet
scalar, which leads to a cosmology with a bounded Hubble expansion rate.

One may ask if a set of metric f(R) theories which include ACM as an
homogeneous and isotropic solution exists.
The answer is that a bi-parametric family of f(R) actions which lead to
an ACM solution exists. In the following section we will derive them
and we will discuss some examples. The derivation follows the same
steps of modified f(R)-gravity reconstruction from any FRW cosmology
\cite{Odin4}.

We start from the action
\be
 S=\frac 1{16 \pi G_N}\int d^4 x\sqrt{-g}[f(R) + 16\pi G L_m]\, ,
\ee
which leads to the ``generalized first Friedman equation''
\begin{eqnarray}
\begin{array}{c}
 -18(H \ddot H +4 H^2 \dot H)f''(R)\\
-3(\dot H +H^2)f'(R)
-\frac 12 f(R)=\left( 8\pi G_N \right) \rho
\end{array}\\
R=-6(\dot{H}+2H^2)\, .
\end{eqnarray}
If the energy density of the universe is mainly due to matter (as in
the late accelerated expansion), we can use (\ref{matter}) and
(\ref{rho_g}) to express $\rho$, $R$, $\dH$ and $\ddH$ as functions of
$H$. Thus we get
\begin{eqnarray}
\begin{array}{c}
  [H^3 g g'^2-3H^2 g^2 g'+3H^3 g^2 g'']f''(R)\\
+\frac{[3H g g'^2-H^2 g'^3]}{18}f'(R)-\frac{g'^3}{108}f(R)=\frac{g g'^3}{18}
\end{array}
\label{difR}\\
R(H) = 18 H g(H)/g'(H)-12 H^2 \label{RH}\, .
\end{eqnarray}
$H(R)$ can be obtained from (\ref{RH}) and set into (\ref{difR}); then
we get an inhomogeneous second order linear differential equation with
non-constant coefficients. Therefore, there will always be a
bi-parametric family of $f(R)$ actions which present ACM as their
homogeneous and isotropic solution. The difference between these
actions will appear in the behavior of perturbations, which is not
fixed by (\ref{rho_g}). Some of these actions are particularly easy to
solve. If $g(H)=H^2$ then $R=-3H^2$ and the differential equation
becomes
\begin{equation}
6 R^2 f''(R) -R f'(R)-f(R) +2 R=0\, .
\end{equation}
Its general solution is
\be
f(R)=R+c_1 R^{\frac 1{12}(7-\sqrt{73})}+c_2 R^{\frac 1{12}(7+\sqrt{73})}\, ,
\ee
which will give (\ref{rho_g}) as First Friedman equation as long as
the radiation energy density can be neglected. In general the
differential equation (\ref{difR}) will not be solvable
analytically, and only approximate solutions can be found as power
series around a certain singular point $R_0$ (a value of $R$ such that
$H(R)$ cancels out the coefficient of $f''(R)$ in (\ref{difR})). These
solutions will be of the form $f(R)=f_p(R) +c_1 f_1(R) +c_2 f_2(R)$
with
\be
\frac{f_i(R)}{R_0}=\sum_{m=0}^\infty a_m^{(i)} (\frac{R-R_0}{R_0})^{s_i+m} \, ,
\ee
where $i=p,1,2$, $a_0^{(1)}=a_0^{(2)}=1$ and the series will converge
inside a certain radius of convergence. An interesting choice of $R_0$
is $R_0 = -12 H_-^2 \equiv R_-$, which is the value of $R$ at
$H_-$. One can also find the approximate solution of the differential
equation for $\mid R\mid\gg\mid R_-\mid$, which will be of the form
\be
\frac{f_i(R)}{R}=\sum_{m=0}^\infty a_m^{(i)} (\frac{R_-}{R})^{s_i+m} \label{fR-}\, .
\ee
One can in principle assume that the action (\ref{fR-}) could be
considered as valid also in the region in which radiation begins to
dominate, but this action reproduces (\ref{rho_g}) only if matter
dominates. However, some of the new terms appearing in (\ref{fR-})
will be negligible against $R$ when radiation begins to dominate. The
others can be canceled out by setting to zero the appropriate integration
constant and therefore the Cosmological Standard Model will be
recovered as a good approximation when radiation begins to dominate.

 $f(R)$-theories are not the only modified gravity theories studied in
the literature; $f(G)$-theories \cite{GB} are also a popular field of research.
In these theories, the Einstein-Hilbert action is supplemented by a
function of the Gauss-Bonnet scalar
$G=R^2-4R_{\mu\nu}R^{\mu\nu}+R_{\mu\nu\rho\sigma}R^{\mu\nu\rho\sigma}$
(not to be confused with the gravitational coupling $G_N$).The same
approach can be used to answer what $f(G)$ actions are able to
reproduce ACM homogeneous evolution, once more in analogy with the
modified f(G)-gravity reconstruction of a FRW cosmology
\cite{Odin5}. The form of the action will be
\begin{equation}
 S=\frac 1{16 \pi G_N}\int d^4 x\sqrt{-g}[R+f(G) + 16\pi G_N L_m]\, ,
\end{equation}
from which we derive the modified Friedman equation
\begin{equation}
 \left(8\pi G_N\right) \rho = 3 H^2-\frac 12 G f'(G)+\frac 12
 f(G)+12f''(G)\dot{G}H^3 \, .
\end{equation}
Following the same procedure as in the case of $f(R)$-theories we
arrive to the differential equation
\begin{eqnarray}
\begin{array}{c}
  3 g(H) = 3 H^2-\frac 12 G f'(G)+\frac 12 f(G)\\
+\frac{864 H^6 g}{g'^3}[9 g g'-H g'^2-3 H g g''] f''(G)
\end{array}
\label{difG}\\
G=24 H^2(H^2-3H g/g')\label{GH}\, .
\end{eqnarray}
For a given $g(H)$, one can use (\ref{GH}) to get $H(G)$, then set it
into (\ref{difG}) and solve the second order linear differential
equation. As in the previous case, there will be a bi-parametric family
of solutions for $f(G)$ which will have (\ref{rho_g}) as their
homogeneous isotropic solution as long as matter dominates. The
parameters will need to be fixed in order to make the contribution of
undesired terms in the Friedman equation to be negligible when
radiation dominates. Again $g(H)=H^2$ is an example which can be
solved analytically. The differential equation turns to be
\be
12 G^2 f''(G)-G f'(G)+f(G)=0\, ,
\ee
which has the solution
\be
f(G)=c_1 G + c_2 G^{1/12}\, .
\ee
In most cases this procedure will not admit an analytical solution
and an approximate solution will need to be found numerically.

A similar discussion can be made to explain the early accelerated
expansion (inflation) as a result of an $f(R)$ or $f(G)$ action,
taking into account that in this case the dominant contribution to the
energy density is radiation instead of matter. The analogue to
(\ref{fR-}) will be now a solution of the form
\be
\frac{f_i(R)}{R}=\sum_{n=0}^\infty a_n^{(i)} (\frac{R}{R_+})^{r_i+n} \label{fR+}\, ,
\ee
for $\vert R \vert \ll \vert R_+\vert$ with $R_+ = -12 H_+^2$.
Terms in the action which dominate over $R$ when $R$ becomes small
enough should be eliminated in order to recover the Standard
Cosmological model before matter starts to dominate.

Both accelerated expansions can be described together in the
homogeneous limit by an $f(R)$ action with terms
$(\frac{R_-}{R})^{s_i+m}$ with $s_i>0$ coming from (\ref{fR-}) and
terms $(\frac{R}{R_+})^{r_i+n}$ with $r_i>0$ coming from
(\ref{fR+}). The action will contain a term, the Einstein-Hilbert
action $R$, which will be dominant for $H_-\ll H\ll H_+$ including the
period when matter and radiation have comparable energy densities.

\section{Alternative Descriptions of the ACM}

In general, by virtue of the gravitational field equations, it is
always possible to convert a modification in the gravitational term of
the action to a modification in the matter content of the universe. In
particular, at the homogeneous level, it is possible to convert a
generalized first Friedman equation of the type (\ref{rho_g}) to an
equation in which, apart from the usual matter term, there is a dark
energy component with an unusual equation of state $p=p(\rho)$ and in
which General Relativity is not modified.

The trivial procedure is the following. Assume that the source of the
gravitational field in the modified gravity theory behaves as
$p_d=\omega \rho_d$ (matter or radiation). This is the case when the
modification of gravity is relevant. We can then define a dark energy
or effective gravitational energy density as

\be
 \rho_g =\frac{3}{8\pi G_N} \left( H^2-g(H)\right)  \label{dedensity}\, .
\ee
The continuity equation (\ref{cont_eq}) allows to define a pressure
for this dark energy component as $p_g=-\rho_g-\dot{\rho}_g/3 H$. The
equation of state of the dominant content of the universe leads to
\be
\dH=-3(1+\omega)\frac{H g(H)}{g'(H)} \label{dH}\, ,
\ee
which can be used to express $\dot{\rho}_g$ as a function of $H$, and
one gets the final expression for $p_g$,
\be
p_g=\frac{-3}{8\pi G_N}\left(
 H^2-2(1+\omega)H\frac{g(H)}{g'(H)}+\omega g(H)\right)
 \label{depressure}\, .
\ee
Then, for any given $g(H)$ we can use (\ref{dedensity}) to get
$H=H(\rho_g)$ and substitute in (\ref{depressure}) to get an
expression of $p_g(\rho_g;H_\pm,\omega)$ which can be interpreted as
the equation of state of a dark energy component. In the simple case
of a matter dominated universe with ACM, $\alpha_-=1$ gives obviously
a dark energy component verifying $p_g=-\rho_g$. Another simple
example is $\alpha_-=2$, for which
\be
\left(\frac{8 \pi G_N}{3 H_-^2} \right) p_g=\frac{2}{\frac{8 \pi
    G_N}{3 H_-^2}\rho_g -3}\, .
\ee

The inverse procedure is also straightforward. Suppose we have a
universe filled with a standard component $p_d=\omega\rho_d$ and a
dark energy fluid $p_g=p_g(\rho_g)$. Using the continuity equation of
both fluids one can express their energy densities as a function of
the scale factor $a$ and then use this relation to express $\rho_g$ as
a function of $\rho_d$. Then the Friedman equation reads
\be
H^2=\frac{8\pi G_N}{3}(\rho_d+\rho_g(\rho_d))\, ,
\ee
which, solving for $\rho_d$, is trivially equivalent to (\ref{rho_g}).
This argument could be applied to reformulate any cosmological model
based on a modification of the equation of state of the dark energy
component \cite{Odin3} as a generalized first Friedman equation
(\ref{rho_g}).

Until now we have considered descriptions in which there is an exotic
constituent of the universe besides a standard component (pressureless
matter or radiation). In these cases, pressureless matter includes both
baryons and Dark Matter. However, there are also descriptions in which
Dark Matter is unified with Dark Energy in a single constituent of the
universe. One of these examples is the Chaplygin gas $p_g=-1/\rho_g$
\cite{Chaplygin, Fabris}. The use of the continuity equation leads to
\be
\rho_g=\sqrt{A+B a^{-6}}\, ,
\ee
where $A$ and $B$ are integration constants. The previous method can
be used to find the generalized first Friedman equation for the baryon
density in this model,
\be
\frac{8\pi
  G_N}{3}\rho_b=\frac{H^2}{k}\left(\sqrt{1+k(1-\frac{H_-^4}{H^4}})-1
\right) \label{Chap}\, ,
\ee
where $H_-=\sqrt{\frac{8\pi G}{3}}A^{1/4}$, $k=B \rho_{b0}^{-2}-1$,
and $\rho_{b0}$ is the present value of $\rho_b$. In the $H\gg H_-$
limit this model can be interpreted as a universe filled with baryons
and dark matter or as a universe filled with baryons and with a higher
effective value of $G_N$. Equation (\ref{Chap}) does not fulfill
(\ref{hpm}) because it describes the behavior of just baryon
density. In the $H\gtrsim H_-$ limit, the model can be interpreted as
a universe filled with baryons and a cosmological constant or as an
ACM model with $\alpha_-=1$ and filled only with baryons.

A similar example in the early period of accelerated expansion would
be a universe filled with a fluid which behaves as
a fluid of ultra-relativistic particles if the energy density is low
enough but whose density has an upper bound
\be
\rho_X=\frac\sigma{a^4+C}\, .
\ee
This dependence for the energy density in the scale factor follows
from the equation of state
\be
p_X=\frac 13 \rho_X-\frac {4 C}{3 \sigma}\rho_X^2 \, .
\ee
This model turns out to be equivalent to a universe filled with
radiation following eq. (\ref{alpha+}) with $\alpha_+=1$ and
$H_+^2=(\frac{8 \pi G_N}{3})\frac{\sigma}{C}$.

Another equivalent description would be to consider that the universe
is also filled with some self interacting scalar field $\varphi$ which
accounts for the discrepancy between the standard energy-momentum and
GR Einstein tensors. Given an arbitrary modified Friedman equation
(\ref{rho_g}) a potential $V(\varphi)$ can be found such that the
cosmologies described by both models are the same. The procedure is
similar to the one used to reconstruct a potential from a
given cosmology \cite{Odin7}. From the point of
view of the scalar field, the cosmology is defined by a set of three
coupled differential equations
\begin{eqnarray}
 \ddot{\varphi}+3 H \dot{\varphi}+V'(\varphi)&=0 \label{klein}\\
\frac{8\pi G_N}{3}(\rho_d+\frac{\dot{\varphi}^2}2 +V(\varphi))&=H^2
\label{quint}\\ -3(1+\omega)H\rho_d&=\dot{\rho_d} \label{contw}\, .
\end{eqnarray}
The solution of these equations for a certain $V(\varphi)$ will give
$H(t)$ and $\rho_d(t)$, and therefore $\rho_d(H)$ which is $g(H)$ up
to a factor $\frac{8 \pi G_N}{3}$. In this way one finds the
generalized first Friedman equation associated with the introduction
of a self interacting scalar field.
Alternatively, given a function $g(H)$ in a generalized first Friedman
equation (\ref{rho_g}) for the density $\rho_d$, one can find a scalar
field theory leading to the same cosmology in the homogeneous
limit. By considering the time derivative of (\ref{quint}) and using
(\ref{klein}) and (\ref{contw}) one gets
\be
\dot{\varphi}^2=-(1+\omega)\rho_d-\frac{\dH}{4\pi G_N} \label{dtvarphi}\, ,
\ee
where we can use (\ref{dH}) and (\ref{rho_g}) to get $\dot{\varphi}^2$
as a function of $H$. Setting this on (\ref{quint}) we get
$V(\varphi)$ as a function of $H$. On the other hand,
$\varphi'(H)=\dot{\varphi}/\dH$, so
\begin{equation}
\frac{8\pi G_N}{3}(\varphi'(H))^2=\frac{(2H
 -g'(H))g'(H)}{9(1+\omega) H^2 g(H)}
\label{phi}
\end{equation}
and
\begin{equation}
\frac{8\pi G_N}{3}V(\varphi(H))=H^2-g(H)+\frac{(1+\omega)(2H-g'(H))g(H)}{2
   g'(H)}\, .
\label{V(phi)}
\end{equation}
If $g(H)$ is such that the rhs of (\ref{phi}) is positive definite, it
can be solved and the solution $\varphi(H)$ inverted and substituted
into (\ref{V(phi)}) in order to get $V(\varphi)$. This is the case of
a function of the type (\ref{alpha-}) with $\alpha_->1$. For the
particular case of an ACM expansion with $\alpha_-=1$, the solution is
a flat potential $V=V_0$ and a constant value of $\varphi=\varphi_0$.

If $g(H)$ is such that the rhs of (\ref{phi}) is negative definite, as
it happens in (\ref{alpha-}) with $\alpha_-<1$ or in (\ref{alpha+}),
the problem can be solved by changing the sign of the kinetic term in
(\ref{quint}). The result is a phantom quintessence in the case of
(\ref{alpha-})  with $\alpha_-<1$. The case (\ref{alpha+}) is more
complicated because the energy density of the associated inflaton
turns out to be negative. Moreover, it is of the same order as the
energy density of ultra-relativistic particles during the whole period
of accelerated expansion. Therefore, it seems that this model is
inequivalent to other inflation scenarios previously studied.

In summary, there are many equivalent ways to describe the discrepancy
between observed matter content of the universe and Einstein's General
Relativity. At the homogeneous level, it is trivial to find relations
among them. The possibility to establish the equivalence of different
descriptions is not a peculiarity of the description of this discrepancy
in terms of a generalized first Friedman equation (\ref{rho_g}). The
same relations can be found among  generalized equations of state,
scalar-tensor theories and f(R) modified gravity \cite{Odin6}.

\section{Summary and discussion}
It may be interesting to go beyond $\Lambda$CDM in the description of
the history of the universe in order to identify the origin of the
two periods of accelerated expansion. We have proposed to use the
expression of the energy density as a function of the Hubble parameter
as the best candidate to describe the history of the universe. In this
context the late time period of accelerated expansion and the early
time inflation period can be easily parametrized.

We have considered a simple modification of the cosmological equations
characterized by the appearance of an upper and a lower bound on the
Hubble expansion rate. A better fit of the experimental data can be obtained
with this extended cosmological model as compared with the
$\Lambda$CDM fit. Once more precise data are available, it will be
possible to identify the behavior of the energy density as a function
of the Hubble parameter and then look for a theoretical derivation of
such behavior.

We plan to continue with a systematic analysis of different
alternatives incorporating the main features of the example considered
in this work. Although the details of the departures from the standard
cosmology can change, we expect a general pattern of the effects induced
by the presence of the two bounds on $H$. We also plan to go further,
considering the evolution of inhomogeneities looking for new consequences
of the bounds on $H$.

The discussion presented in this work, which is based on a new description
of the periods of accelerated expansion of the universe, can open a new way
to explore either modifications of the theory of gravity or new components
in the universe homogeneous fluid. Lacking theoretical criteria to select
among the possible ways to go beyond $\Lambda$CDM, we think that the
phenomenological approach proposed in this work is justified.

We are grateful to Paola Arias and Justo L\'opez-Sarri\'on for
discussions in the first stages of this work. We also acknowledge
discussions with Julio Fabris, Antonio Segu\'{\i}, Roberto Empar\'an,
Sergei Odintsov, Aurelio Grillo and Fernando M\'endez.
This work has been partially supported by CICYT (grant
FPA2006-02315) and DGIID-DGA (grant2008-E24/2). J.I. acknowledges a
FPU grant from MEC.


\end{document}